\documentclass[12pt,a4paper]{article}
\usepackage[left=15mm,right=15mm,top=15mm,bottom=15mm]{geometry}
\usepackage{graphicx}
\usepackage{multirow}
\usepackage{float}
\usepackage{xcolor}
\usepackage{color}
\usepackage{hyperref}
\usepackage{booktabs} % tables stuff
\usepackage{algorithm}
\usepackage{algpseudocode}
\usepackage{amsmath} % formula de la distancia coseno
\usepackage{longtable} % tabla mas larga de una pagina
\usepackage{caption}
\usepackage{subcaption} % para el comando \subcaption
\usepackage{colortbl}
\usepackage[backend=biber]{biblatex}
\title{Shared quasispecies architecture in experimental and natural RNA virus populations}
\author{
\normalsize    Samuel Martínez-Alcalá$^{1,2,*}$, 
Iker Atienza-Diez$^{1,2,*}$, Pilar Somovilla$^{3}$, \\
\normalsize Brenda Martínez-González$^{4,5}$, Celia Perales$^{4,5}$, Luis F. Seoane$^{6}$, \\
\normalsize
Ester Lázaro$^{3}$, Susanna Manrubia$^{1,2}$}
\date{
    \normalsize
$^1$ Museo Nacional de Ciencias Naturales (CSIC), c/ José Gutiérrez Abascal 2, 28006 Madrid, Spain \\
$^2$ Grupo Interdisciplinar de Sistemas Complejos (GISC), 28911 Madrid, Spain \\
$^3$ Centro de Astrobiología (CAB), CSIC-INTA, Carretera de Ajalvir Km 4, 28850 Torrejón de Ardoz, Madrid \\
$^4$ Centro de Biología Molecular ``Severo Ochoa'' (CSIC-UAM), Campus de Cantoblanco, 28049 Madrid, Spain \\  
$^5$ Instituto de Investigación Sanitaria-Fundación Jiménez Díaz University Hospital (UAM), 28040 Madrid, Spain \\
$^6$ Institut de Biologia Evolutiva (UPF-CSIC), BIT Lab, Barcelona 08003, Spain \\ 
$^*$ Equal contribution \\
\today
}  

\begin{document}

\maketitle

\begin{abstract}
RNA viruses form genetically diverse populations structured as mutant spectra, or quasispecies, whose internal organization influences their evolutionary and adaptive dynamics. While genetic diversity has been extensively characterized, the structural organization of viral populations in sequence space remains less explored. Here, we compare genotype network architectures in two RNA viruses with markedly different evolutionary contexts: bacteriophage $Q\beta$ evolving in controlled laboratory conditions and SARS-CoV-2 evolving within infected human hosts. Using deep sequencing data, we reconstruct the genotype network of mutationally coupled variants within viral populations and analyze their topological properties. Despite large differences in genome size, mutation rate, and ecological setting, both viruses exhibit a common organization: a highly abundant central haplotype surrounded by layers of variants of diminishing abundance as Hamming distance to the central haplotype increases. All reconstructed networks share qualitative and quantitative topological features, displaying a hierarchical structure. The robust organization of both populations under multiple conditions suggests that RNA viruses may share a common genotype network architecture governed by fundamental properties of sequence space and the generic mechanisms of replication and mutation. Genotype networks provide a unifying framework to describe viral population structure beyond conventional diversity measures and, by revealing how local constraints shape mutational search, offers  insights into the predictability of viral evolution.
%209 words; max 350
\end{abstract}

\section{Introduction}

RNA viruses replicate with high error rates, giving rise to genetically heterogeneous populations composed of closely related genomes. These populations are commonly described as viral quasispecies \cite{andino:2015}, a concept originally formulated in theoretical studies of self-replicating molecules \cite{eigen:1971} and later extended to RNA virus evolution \cite{domingo:1978}. In this framework, viral populations are viewed as dynamic distributions of mutants shaped by the interplay between mutation and selection. Such mutant spectra constitute the raw material for rapid adaptation to environmental challenges, including new hosts, immune pressures, and antiviral therapies \cite{mattenberger:2021,somovilla:2022,haddox:2026}.
%, and they play a central role in viral pathogenicity \cite{vignuzzi:2006}, persistence and transmission.
  
Early experimental and theoretical work on RNA viruses consistently showed that viral populations are not dominated by a single genotype. Rather, they are composed of a structured distribution of mutants around a most abundant sequence. In bacteriophage Q$\beta$, classic studies demonstrated that mutation continuously generates single-step variants, such that most genomes differ from the consensus at only one or a few positions, and where the frequency of genotypes declines with mutational distance from the master sequence \cite{batschelet:1976,domingo:1978}. Similar conclusions were reached for vesicular stomatitis virus, where populations were described as mutant ``clouds'' tightly clustered around a central sequence, with progressively fewer high-distance variants due to the action of purifying selection \cite{holland:1982}. This is a rough but ubiquitous pattern subsequently observed in other RNA viruses, such as hepatitis C \cite{martell:1992}, human immunodeficiency virus \cite{henn:2012} or SARS-CoV-2 \cite{delgado:2024}. Together, these observations support a general view in which viral mutant spectra are strongly concentrated around a central genotype, with abundance decreasing as mutational distance increases due to the combined effects of mutation and selection \cite{lauring:2010}.

Empirical work on RNA viruses has often focused on quantifying genetic diversity within populations \cite{sanjuan:2021} and in exploring how diversity relates to viral fitness, pathogenesis or adaptation \cite{lauring:2010,borderia:2011}. High-throughput sequencing has enabled increasingly detailed descriptions  
of mutant spectra across a wide range of viruses \cite{posadacespedes:2017,perezlosada:2020}, including human pathogens such as hepatitis C virus \cite{quer:2017} and SARS-CoV-2 \cite{lythgoe:2021}. Analyses of SARS-CoV-2 mutant spectra in infected patients have shown that intra-host genetic complexity correlates with clinical outcomes \cite{martinezgonzalez:2022_MS} and may represent an epidemiologically relevant trait \cite{martinezgonzalez:2022,martinezgonzalez:2025}. 
%However, quantifying diversity alone provides only partial information about the organization of viral populations.
Notably, populations with similar levels of diversity can still exhibit distinct patterns in the organization of their mutant clouds. The internal architecture provides complementary information on the likelihood that minority variants persist in the population, on the range of mutations actively explored by the quasispecies as a whole, and also on the evolutionary trajectories available to the virus under changing selective pressures.

Despite previous studies on the ``cloud'' of mutants surrounding the most abundant sequence, the detailed structure of the quasispecies, namely, how individual haplotypes are mutationally related and how their abundances are distributed across the underlying genotype space, is a relevant but still poorly explored feature. 
This internal organization can be studied through the reconstruction of genotype networks, defined as graphs of mutationally connected variants in which haplotypes differing by a single position are linked. Genotype networks derived from consensus haplotypes have uncovered the extent of convergent evolution and positive selection in influenza A \cite{aguilar:2018}, as well as correlations between viral diversity and local network structure \cite{williams:2022}. However, 
studies addressing the characterization of genotype networks for intra-host viral diversity are scarce so far.    
%From an evolutionary perspective, this structure determines the extent to which populations explore sequence space, their mutational robustness, and their potential for future adaptation. 

A recent work on the RNA bacteriophage Q$\beta$ has analyzed the structure of a set of populations evolved through serial passages at different temperatures that impose distinct selective pressures on the virus. Genotype networks reconstructed from deep-sequencing data revealed the existence of hierarchical networks centered around the most abundant haplotype, with variants arranged in concentric layers corresponding to increasing Hamming distance from this central sequence \cite{seoane:2026}. This structure determines the extent to which populations explore sequence space. Remarkably, this hierarchical organization persists across evolutionary passages and environmental conditions, suggesting that it may arise from simple and general processes. Whether such a structure is generalizable to different RNA viruses, subject to different selective pressures or evolving in natural infections remains an open question.

In this contribution, we perform a systematic comparison between bacteriophage Q$\beta$ and SARS-CoV-2, two positive-sense RNA viruses with markedly different biological and evolutionary contexts. The SARS-CoV-2 genome ($\approx 30$ kb) is about an order of magnitude larger than that of Q$\beta$ and encodes a proofreading exonuclease \cite{moeller:2022} that reduces the effective mutation rate relative to many RNA viruses \cite{symons:2025}. Indeed, analyses of biological clones show that viable SARS-CoV-2 genomes accumulate on average several-fold fewer mutations than other RNA viruses such as foot-and-mouth disease virus \cite{deavila:2025}. In addition, SARS-CoV-2 populations evolve within infected hosts under immune pressure and are transmitted through epidemiological chains characterized by severe population bottlenecks \cite{lythgoe:2021} followed by periods of intra-host replication and diversification.
 
%Despite these differences, 
%Both viruses generate mutant spectra that can be studied with deep sequencing approaches. 
%In SARS-CoV-2, analyses of intrapatient populations have alreay revealed extensive genetic heterogeneity and complex frequency distributions of mutations, including distinctive patterns such as a relative deficit of variants at intermediate frequencies compared with other RNA viruses (Martínez-González et al., 2022b). Such observations raise the question here is whether the topological organization of viral mutant spectra might also display common features across viruses with very different genome sizes, replication mechanisms and ecological contexts.

Using deep sequencing data, we reconstruct and compare the genotype network structure of Q$\beta$ populations evolving {\it in vitro} with that of SARS-CoV-2 populations sampled from infected patients. Our results reveal striking similarities between the two systems. Despite the profound differences in their evolutionary histories and replication environments, their genotype networks display the same hierarchical architecture, characterized by a highly connected central haplotype surrounded by layers of increasingly rare mutants. We trace minor quantitative differences in the organization patterns of the two viruses to the coverage of the deep sequencing protocol, which causes variation in the depth of sequence space exploration revealed by our samples. These findings suggest that the structural organization of viral quasispecies may be governed primarily by universal properties of RNA virus replication and sequence space rather than by virus-specific ecological conditions.

\section{Materials and Methods}
\label{sec:data}

\subsection{Q$\beta$ phage data}
Q$\beta$ phage has a short RNA genome (4217 nt) that encodes four proteins (see Fig.\ \ref{fig:genomes}a). It infects the bacterium {\it Escherichia coli} using the conjugative F pilus as a receptor. Its optimal replication temperature is $37\> ^{\circ}$C, although it can readily adapt to replicate between $30\> ^{\circ}$C and $43\> ^{\circ}$C \cite{arribas:2014,somovilla:2019,arribas:2021,somovilla:2022}. 

For this study, we consider representative populations from various experiments in which the phage was evolved at different constant temperatures ($30$, $37$, $40$ and $43\> ^{\circ}$C), starting from a common ancestral population (see Supplementary Material for an exception). Experiments proceeded for up to 60 passages, and population samples from different passages were deep sequenced (see \cite{seoane:2026} for a detailed description of the experimental procedures and deep-sequencing protocols).
%Some topological features of populations at passage 60 of these experiments were analysed in \cite{seoane:2026}. 
The evolution temperatures were chosen based on previous laboratory experience \cite{arribas:2014,somovilla:2019,arribas:2021,somovilla:2022}, taking into account that extreme temperatures impose strong selective pressures on both the virus and the host. At $30\> ^{\circ}$C, host metabolic processes are slowed down, whereas at $43\> ^{\circ}$C the heat shock response is fully activated in {\it E. coli}. In Q$\beta$, temperature has a strong impact on replication dynamics by affecting RNA stability, protein folding, and the efficiency of viral assembly.

\subsection{SARS-CoV-2 data}

SARS-CoV-2 has a non-segmented, large positive-sense single-stranded RNA genome of approximately 29.9 kb, comprising 14 open reading frames (ORFs) and encoding 27 proteins (16 non-structural, 4 structural, and 7 accessory proteins), see Fig.\ \ref{fig:genomes}b. For this study, we analyze samples from patients admitted to Fundación Jiménez Díaz Hospital (Madrid, Spain) during the first COVID-19 outbreak in Spain, within a short time window early in the pandemic and during the transition between clades 19 and 20. All patients were confirmed to be positive for SARS-CoV-2 \cite{martinezgonzalez:2022_MS}.

Patients were classified based on COVID-19-associated clinical parameters, including the need for hospitalization, requirement for mechanical ventilation, admission to the intensive care unit, and death attributed to COVID-19. According to these criteria, cases were categorized as mild (no hospital or ICU admission), moderate (hospitalization without ICU admission), and severe (ICU admission, all resulting in death). In the present analysis, we focus on a subset of patients, including two representative cases from each clinical category (mild, moderate, and severe).

\subsection{Data curation}

From the available datasets for the two viruses, a subset of samples was selected for detailed comparative analysis. Specifically, representative samples were chosen for each clinical category in SARS-CoV-2 (mild, moderate, and severe), and one intermediate passage was selected for each temperature condition in $Q\beta$. Two different amplicons were analysed: in the main text we show and discuss results for amplicon 1 in both viruses; results for amplicon 2 are provided as Supplementary Information.

Amplicons of the Q$\beta$ genome were deep sequenced in populations at passage 30. The first fragment spans nucleotides 1060 to 1330 (amplicon 1); positions 1060 to 1320 code for a fragment of the A2 maturation-lysis protein, 1321 to 1323 correspond to a termination codon, and 1324 to 1330 are non-coding positions. The second fragment (amplicon 2) spans nucleotides 2145 to 2473, coding for fragments of the A1 coat protein (2145--2330), a non-coding region (2331--2351), and the replicase (2352--2473). 

\begin{figure}[t]
    \begin{center}
        \includegraphics[width=\textwidth]{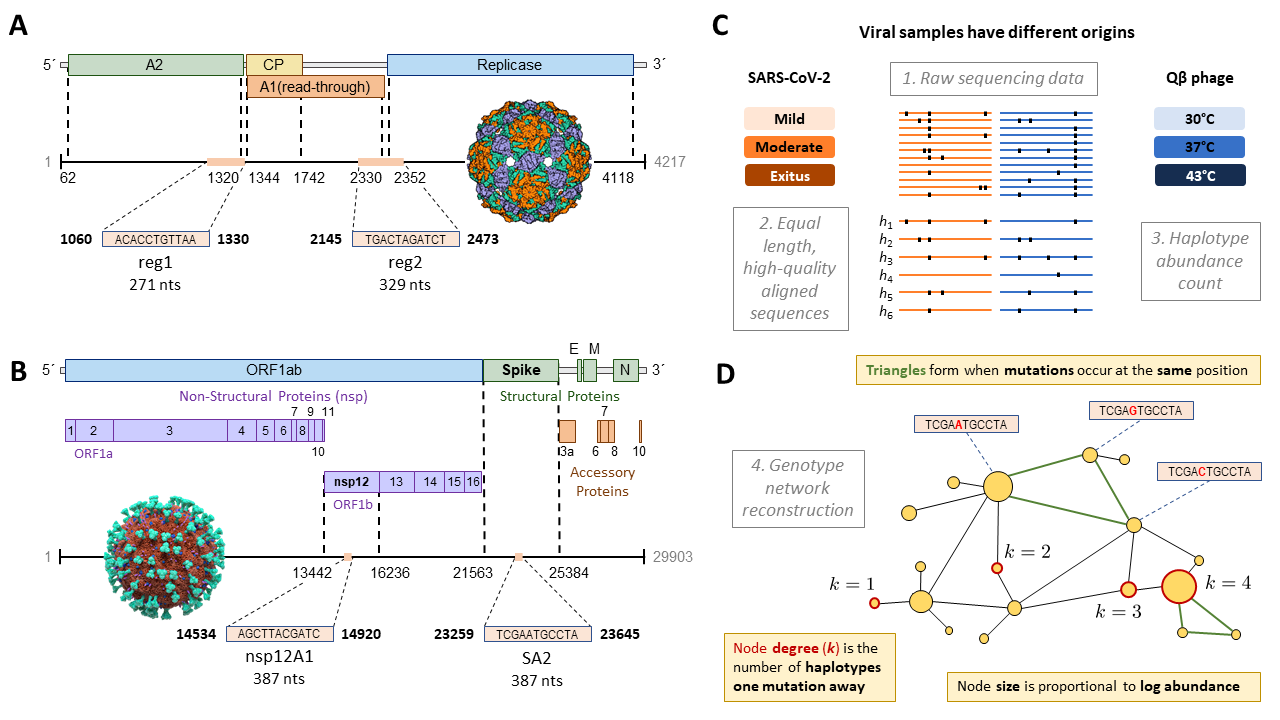}
        \caption{\textbf{Schematic of viral genomes and genotype network reconstruction.} 
        (a) Genome architecture of the Q$\beta$ phage and (b) SARS-CoV-2. The positions and relative sizes of the two amplicons used for genotype network reconstruction are indicated. 
        (c) Summary of data origin and filtering pipeline, with main procedures numbered. 
        (d) Illustration of genotype network reconstruction and key topological quantities.}
        \label{fig:genomes}
    \end{center}
\end{figure}

In the case of SARS-CoV-2, the two selected amplicons were deep sequenced in samples obtained from infected patients. The first fragment spans nucleotides 14534 to 14920 and codes for part of the non-structural protein nsp12, corresponding to the viral RNA-dependent RNA polymerase (RdRp). The second fragment spans nucleotides 23259 to 23645 and codes for a region of the structural Spike protein. Figure~\ref{fig:genomes}(a,b) summarizes the viral genomes and illustrates the positions of the analyzed amplicons. FASTQ files derived from SARS-CoV-2 samples within the patient cohort are publicly available in the European Nucleotide Archive (ENA) repository under project ID PRJEB48766. 

This study received approval from the Ethics Committee and the Institutional Review Board of Fundación Jiménez Díaz (FJD) Hospital (no. PIC-087-20-FJD).

Raw data curation in both cases consisted of removing low-quality sequences, sequences not aligned to the reference genome, and sequences containing insertions or deletions. This process resulted in a large set of high-quality sequences of equal length, which were subsequently collapsed into $D$ unique sequences (haplotypes) while retaining their abundances, following previously described protocols \cite{villanueva:2022,seoane:2026}. Figure~\ref{fig:genomes}c schematizes the data curation process. 

After curation, each dataset consisted of a set of haplotypes (one per virus, experimental condition or patient, and passage or sample) and their abundances. In the genotype network representation, each haplotype corresponds to a node, and two nodes are connected if they differ by a single nucleotide (Fig.~\ref{fig:genomes}d). In this study, we consider only haplotypes belonging to the largest connected component (LCC) of the genotype network, defined as the largest set of haplotypes connected through single point mutations. This component represents more than 95\% of all sequences and includes only mutationally coupled variants. 
        
In our representation of genotype networks, node size is proportional to the logarithm of abundance, $\ln s(i)$, where $s(i)$ is the number of occurrences of haplotype $i$ in the curated dataset. For convenience, we define the root sequence as the most abundant haplotype in each population. Note that the root haplotype does not necessarily coincide with the reference or consensus genome, labeled WT for short. In the case of SARS-CoV-2, the reference genome is Wuhan, with accession number NCBI NC$_-$045512.2 \cite{martinezgonzalez:2022,martinezgonzalez:2022_MS}, and for Q$\beta$ the reference genome is GenBank AB971354.1 \cite{somovilla:2022,seoane:2026}.

Table~\ref{tab:populations} summarizes the collection of samples analyzed, including the total number of sequences $A$ after curation, the number of sequences $A_{LCC}$ in the LCC, and the number $D$ of haplotypes and links $\mathcal{L}$ in each sample. WT indicates that the root haplotype coincides with the wild type (see Materials and Methods); otherwise, mutations in the root relative to the WT sequence are specified. HD denotes the maximum Hamming distance from the root among sequences in the LCC. Because of the filtering procedure, there is at least one continuous parsimonious mutational pathway (single-nucleotide steps) connecting the root haplotype to any other haplotype in the LCC, even those at distance $HD^{max}$. The last three columns report least-squares fits to selected topological features of the reconstructed genotype networks (see Section~\ref{sec:definitions}). 

\begin{table}
\centering
 \resizebox{\textwidth}{!}{%
\begin{tabular}{cllcccccclll}
\toprule
Virus/Amplicon & Sample & $A$ & $A_{LCC}$ & $D$ & $\cal L$ & $Root$ & $HD^{max}$ & $\alpha$ & $\beta$ & $\delta$ \\
\midrule
Q$\beta$/1 & 30$^\circ$C & 543,458 & 526,875 & 29,463 & 66,203 & WT & 7 & 2.28(1) & 0.57(1) & 0.137(1) \\
Q$\beta$/1 & 37$^\circ$C & 264,633 & 261,066 & 9,755 & 20,595 & WT & 5 & 2.54(2) & 0.76(1) & 0.132(1) \\
Q$\beta$/1 & 43$^\circ$C & 513,817 & 496,749 & 34,418 & 78,477 & G1312A & 7 & 2.21(1) & 0.60(1) & 0.137(2) \\
\midrule
Q$\beta$/2 & 30$^\circ$C & 709,641 & 696,788 & 32,854 & 72,625 & WT & 6 & 2.21(1) & 0.66(1) & 0.142(0) \\
Q$\beta$/2 & 40$^\circ$C & 346,886 & 341,377 & 11,917 & 23,939 & WT & 6 & 2.47(2) & 0.75(1) & 0.138(2) \\
Q$\beta$/2 & 43$^\circ$C & 286,487 & 281,430 & 12,117 & 24,703 & WT & 6 & 2.56(2) & 0.75(1) & 0.141(2) \\
\bottomrule
SC2/1 & Moderate & 147,112 & 139,927 & 12,890 & 25,558  & WT & 4 & 1.87(2) & 0.92(1) & 0.126(2) \\
SC2/1 & Mild & 134,668 & 127,753 & 14,794 & 29,215  & WT & 4 & 1.76(1) & 0.91(1) & 0.129(2) \\
SC2/1 & Severe & 146,847 & 139,949 & 11,201 & 22,279  & WT & 4 & 2.02(3) & 0.93(1) & 0.124(2) \\
\midrule
SC2/2 & Moderate & 172,445 & 161,328 & 16,990 & 33,795  & A23403G & 4 & 1.78(1) & 0.91(1) & 0.129(2) \\
SC2/2 & Mild & 140,672 & 131,577 & 15,749 & 31,091  & A23403G & 4 & 1.77(1) & 0.91(1) & 0.130(2) \\
SC2/2 & Severe & 138,098 & 129,497 & 12,215 & 24,291  & A23403G & 4 & 1.98(3) & 0.92(1) & 0.126(2) \\
\bottomrule
\end{tabular}
 }
\caption{\textbf{Characteristics of viral population samples used in this study.}
$A$, total number of sequences after data curation; $A_{LCC}$, number of sequences in the largest connected component (LCC); $D$, number of haplotypes (nodes) in the LCC; $\mathcal{L}$, number of links in the LCC; Root haplotype: indicates whether it matches the WT or lists mutations relative to WT (see also section 2.3); and $HD^{max}$, maximum Hamming distance from the root within the LCC. SC2 denotes SARS-CoV-2.
%; A1 and A2 refer to amplicons 1 and 2, respectively \footnote{Amplicon 2 for SARS-CoV-2 
In this study, amplicon 2 for SARS-CoV-2 corresponds to amplicon 6 in a previous publication \cite{martinezgonzalez:2022_MS}. The last three columns report fitted values of exponents characterizing topological features of the genotype networks: $\alpha$ for the complementary cumulative degree distribution, $\beta$ for assortativity and $\delta$ for sequence space covering (see Section 4). Figures for amplicon 1 are presented in the main text; results for amplicon 2 are provided as Supplementary Information.}%The RNA of all SARS-CoV-2 populations was extracted using commercial kits; RNA of Q$\beta$ populations was extracted with phenol-chlorophorm methods with the exception of sample at 37$^\circ$C, whose RNA was extracted using commercial kits (see Materials and Methods). 
\label{tab:populations}
\end{table}

%\clearpage
%\newpage

\section{Network reconstruction and visualization}

We systematically compare the structure of genotype networks for Q$\beta$ and SARS-CoV-2 across the populations listed in Table \ref{tab:populations}. Specifically, we examine population structure across different environments and assess potential differences arising from sequencing coverage. For SARS-CoV-2, we analyze one sample per level of disease severity (mild, moderate, and severe), while for $Q\beta$ we consider populations at passage $30$ at each temperature condition. All analyses refer to amplicon 1 (A1). Details of the six populations and their corresponding genotype networks are summarized in Table \ref{tab:populations}.

\begin{figure}[t]
    \begin{center}
        \includegraphics[width=\textwidth]{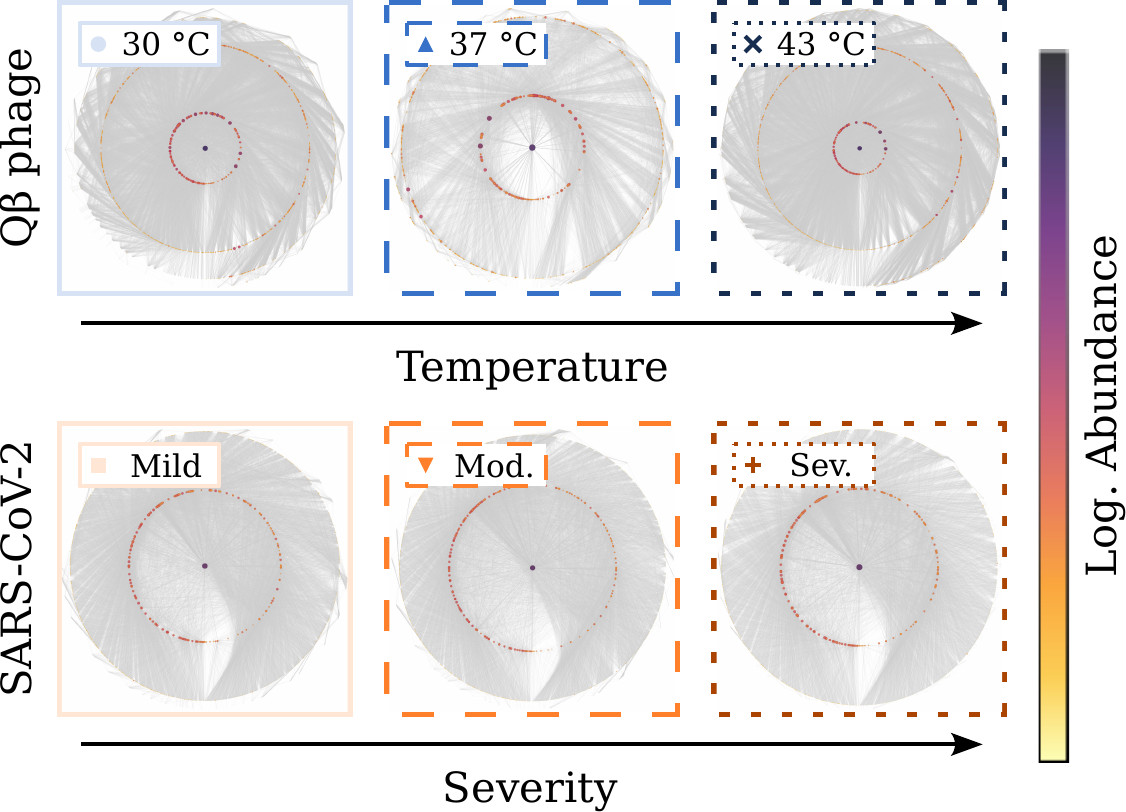}
        \caption{\textbf{Hierarchical representation of genotype networks.} Three representative populations of amplicon 1 for Q$\beta$ phage and SARS-CoV-2 have been chosen to illustrate the hierarchical structure of the LCC genotype network. The most abundant sequence ($Root$) is located in the center of the network, and concentric circles with variants at increasing Hamming distance are represented. The color and line code will be maintained all through this work to facilitate feature comparison.}
        \label{fig:lcc_nets}
    \end{center}
\end{figure}

Figure~\ref{fig:lcc_nets} presents a hierarchical representation of the reconstructed genotype networks. In these representations, genotype networks are constructed by connecting sequences that differ by a single nucleotide, such that the network encodes local mutational accessibility between variants. This framework provides a direct visualization of how viral populations are organized in sequence space beyond simple measures of diversity. In all cases, each node corresponds to a unique nucleotide sequence (haplotype) of the genomic fragment selected for each virus. 

The most abundant haplotype ($Root$) is located at the center of the network and is surrounded by concentric layers of variants at increasing Hamming distance. Thus, proximity to the root reflects the number of mutational steps separating each haplotype from the dominant sequence. Node color and node size both represent the natural logarithm of the abundance. These representations reveal broadly similar structures across all populations. However, the extent of variation around the root differs between the two viruses. SARS-CoV-2 populations contain variants up to three or four mutations away from the root, whereas Q$\beta$ populations reach distances of up to seven mutations ($HD^{max}$; Table~\ref{tab:populations}). This difference may reflect disparities in sequencing coverage between the datasets (column $A$ in Table~\ref{tab:populations}), a possibility that we examine in a later section.

In the next section, we analyze the topology of the reconstructed networks for each virus to determine whether the observed qualitative similarities are also supported by quantitative measures.

\section{Network topology}
\label{sec:definitions}

We focus now on network topology to compare the main features of both viruses, with particular attention to their similarities and differences. Following previous analyses of genotype networks (GN) as complex networks \cite{aguirre:2011,seoane:2026}, we consider three relevant features: the degree distribution $p(k)$, assortativity, and the number of triangles per node. We also measure the following quantities as a function of the Hamming distance $d$ to the root: the total number $D(d)$ of distinct haplotypes, the fraction $q(d)$ of variants present at each distance, an average $dN/dS$ ratio customized for genotype networks, and the degree of exploration of sequence space (sequence space covering, $C(d)$). We begin by defining these quantities.

\subsection{Definitions}

\paragraph{Degree distribution.} The degree $k_i \in [1, 3L]$ of a node $i$ in the LCC is defined as the number of links it has (see Fig.~\ref{fig:genomes}d). This corresponds to the number of haplotypes in the dataset within the 1-mutant neighborhood of the focal haplotype $i$. The degree distribution $p(k)$ represents the probability that a node has degree $k$ in a given sample. A common alternative representation of this probability is the complementary cumulative distribution 
\begin{eqnarray}
P(k' > k) = 1 - \sum_{k' > k} p(k') \, , 
\end{eqnarray}
which yields the fraction of haplotypes with degree $k'$ greater than $k$. The cumulative distribution provides clearer visualization as it averages out noise. This also enables a more robust estimation of distribution parameters. 

\paragraph{Degree assortativity.} Assortativity quantifies how similar two nodes in a network are with respect to a given property, in this case node degree. The average degree $k_{nn}(i)$ of the haplotypes in the 1-mutant neighborhood $nn(i)$ of haplotype $i$ is 
\begin{eqnarray}
k_{nn}(i) = \frac{1}{k_i} \sum_{j \in nn(i)} k_j \, .
\label{eq:assort}
\end{eqnarray}
The assortativity coefficient $r$ is the Pearson correlation coefficient of degree between pairs of linked nodes \cite{newman:2002}. Positive values $r > 0$ indicate that nodes of similar degree tend to be mutually linked (assortative networks).Negative values $r < 0$ reveal a preference for links between nodes of different degree--i.e.\ nodes of lower degree preferentially link to hubs, and vice versa (diassortative networks). Networks in which links are randomly assigned between nodes have values of $k_{nn}(i)$ independent of degree $k$, and $r \simeq 0$.

\paragraph{Triangles.} Triangles in genotype networks are formed when at least three of the possible substitutions at a given sequence position are present in the population sample (see an example in Fig.~\ref{fig:genomes}d). Around a given haplotype $i$, one triangle is formed for each position with two mutants present, while three triangles are formed if all three possible variants at that nucleotide are observed. The total number of triangles $\triangle(i)$ in the 1-mutant neighborhood of haplotype $i$ is the sum of all such contributions along the sequence. 

$\triangle(i)$ is an informative measure of local clustering in genotype networks, as it quantifies deviations from neutral expectations for each haplotype and admits an interpretation in terms of sequence neutrality \cite{seoane:2026}. Specifically, one obtains $\triangle_{rnd}(k) = k^2/(3L)$ if mutations are randomly distributed along the sequence, meaning that all sites have the same degree of neutrality and are equally prone to accept mutations. Significant deviations toward the maximum possible number of triangles, $\triangle_M(k) = k$, indicate variation in site neutrality along the sequence, while deviations toward the minimum possible number of triangles

\begin{equation}
\triangle_m(k) = \left \{
\begin{array}{lc}
0 \, , & 0 \le k \le L; \\
k-L \, , & L \le k \le 2L; \\
2k - 3L \, , & 2L \le k \le 3L;
\end{array}
\right.
\end{equation}
represent situations in which additional mutants at an already mutated position are disfavored. The analytical derivation of $\triangle_{rnd}(k)$, $\triangle_m(k)$, and $\triangle_M(k)$ can be found in \cite{seoane:2026}.

\paragraph{Empirical haplotype diversity.} The total number of distinct haplotypes in the LCC, $D$, of each sample is given in Table~\ref{tab:populations}. This set can be partitioned into subsets according to their Hamming distance $d$ (number of point mutations) from the root sequence. Each subset has size $D(d)$, such that $D = \sum_d D(d)$.

\paragraph{Haplotype diversity explored at distance $d$.} The maximum number ${\cal M}(d)$ of haplotypes at distance $d$ is given by

\begin{equation}
\label{eq:Sd}
{\cal M}(d) = 3^d {L \choose d}
\end{equation}
for a four-letter alphabet \cite{krug:2005,seoane:2026}, where $L$ is the length of the amplicon in each sample. Note that each haplotype has at most $3L$ neighbors. The ratio $D(d)/{\cal M}(d)$ measures the fraction of the total possible diversity ${\cal M}(d)$ explored in each sample at distance $d$.

\paragraph{$dN/dS$ ratio per sequence and averaged for each $d$.} For the coding region of each haplotype, the fraction of non-synonymous versus synonymous mutants in its 1-mutant neighborhood, $w \equiv dN/dS$, is calculated as follows: Given node $i$, we define $S(i)$ as the total number of possible synonymous haplotypes at distance 1, and $\tilde{S}(i)$ as the number of observed synonymous haplotypes at distance 1, regardless of abundance. Similarly, $N(i)$ and $\tilde{N}(i)$ denote respectively the total possible and observed numbers of non-synonymous haplotypes in the 1-mutant neighborhood of $i$ (note that, by definition of the 1-mutant neighborhood, $nn(i) = \tilde{S}(i) + \tilde{N}(i)$). Then
    \begin{eqnarray}
        w(i) &=& { {\tilde{N}(i)/N(i)} \over {\tilde{S}(i)/S(i)} }
        \label{eq:dNdS}
    \end{eqnarray}
is expected to be close to 0 if synonymous (effectively neutral) haplotypes are more frequently observed than haplotypes with non-synonymous mutations around haplotype $i$. The average $dN/dS$ ratio $w_d$ at each distance $d$ is calculated as the mean over all haplotypes at that distance,
    \begin{eqnarray}
        w_d = \frac{1}{D(d)} \sum_{i|d} w(i) \, .
    \end{eqnarray}
This quantity presents a subtle bias towards small values, particularly in low-degree nodes. When few neighbors exists, it might happen that no synonym variants are present, resulting in a $0$ in the denominator of Eq.\ \ref{eq:dNdS} and infinitely large $w(i)$. We must leave these cases out. Below, we show how results from number of triangles (which is unbiased) and $dN/dS$ are consistent for low-degree haplotypes---suggesting the bias is not critical.

\paragraph{Sequence space covering.} To quantify how extensively each population explores sequence space, we define sequence space covering $C(d)$, a measure inspired in shape space covering \cite{gruner:1996b}. Sequence space covering $C(d)$ is defined as the fraction of all possible genotypes at a given Hamming distance $d$ that are actually observed in the population,  
    \begin{eqnarray}
        C(d) = \frac{A(d)}{{\cal M}(d)} \, ,
    \end{eqnarray}
where $A(d)$ stands for sequence abundance at distance $d$.

\subsection{General topological properties}

\begin{figure}[H]
    \begin{center}
        \includegraphics[width=\textwidth]{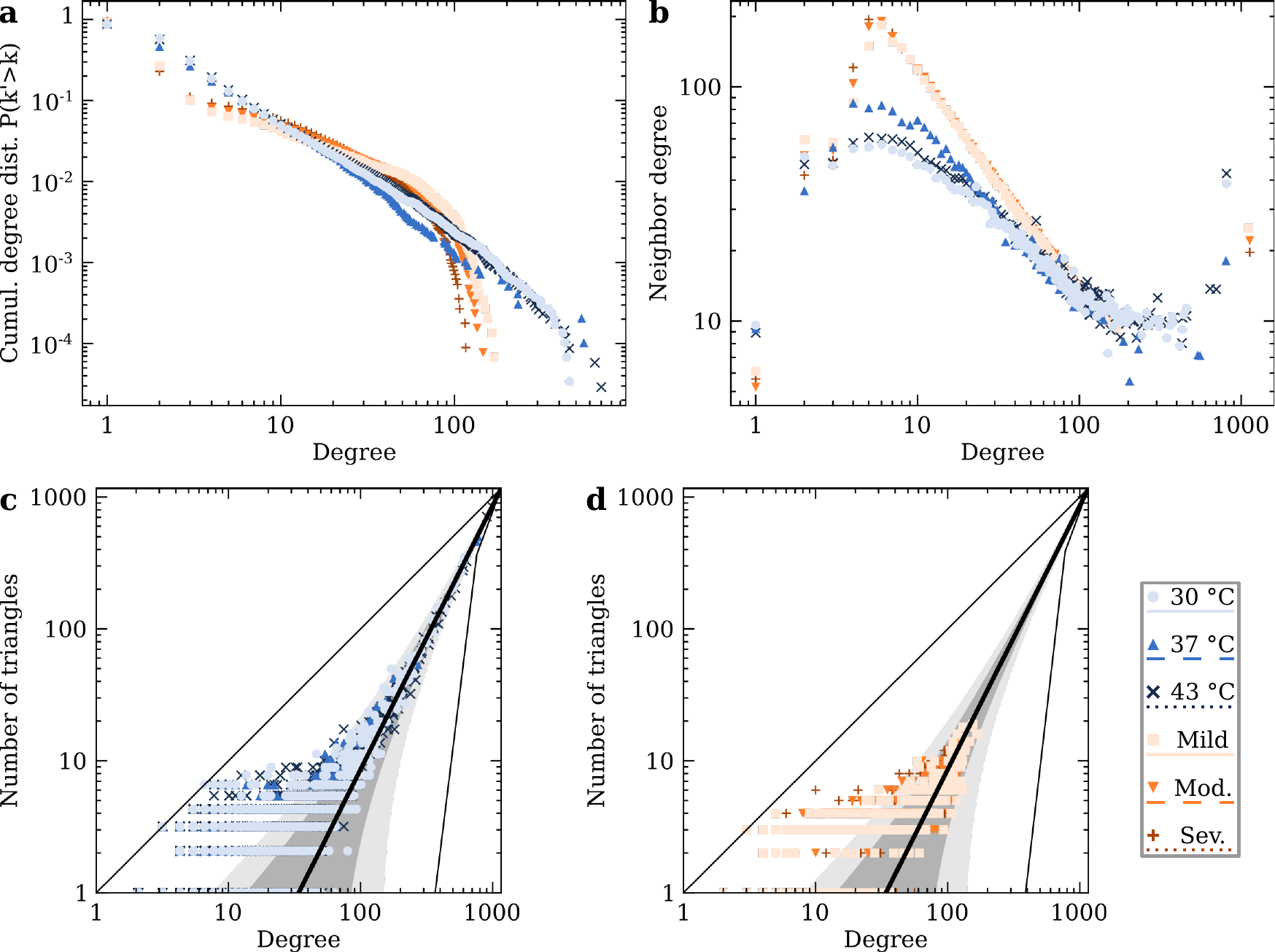}
        \caption{\textbf{Topological structure of the LCC of genotype networks.} Various topological quantities measured for different viral populations are qualitatively similar for Q$\beta$ and SARS-CoV-2. (a) Complementary cumulative distribution function (CCDF), (b) assortativity and (c,d) triangles. In (c,d), black lines indicate the theoretical bounds on the expected number of triangles as a function of node degree (upper and lower lines) and the bold black line represents expected values if mutations are randomly assigned along the haplotype sequence; dark gray and lighter gray regions correspond to one and two standard deviations from average (see \cite{seoane:2026} for the explicit derivation of these quantities). As in previous representations, Q$\beta$ populations are shown in different shades of blue and SARS-CoV-2 in shades of orange, as specified in the legend.}
        \label{fig:macro_pnas}
    \end{center}
\end{figure}

Both the degree distribution $p(k)$ and its complementary cumulative distribution function (CCDF) exhibit fat-tailed behavior, as evidenced by their approximately linear appearance in log-log representations as a function of the degree $k$ (Fig.~\ref{fig:macro_pnas}a). Over a sample-dependent range of $k$, these distributions can be well approximated by a power law of the form $p(k) \propto k^{-\alpha}$, or equivalently $P(k > K) \propto k^{-\alpha + 1}$, where $\alpha$ is the exponent characterizing the degree distribution. This scaling regime spans a broader interval of $k$ values for Q$\beta$ populations, and the corresponding fits are slightly more robust, although SARS-CoV-2 CCDFs also display clear fat-tailed behavior. In the latter case, however, the sampled networks do not reach comparably high maximum degrees, consistent with differences in sampling depth, with SARS-CoV-2 datasets comprising on the order of $10^5$ sequences, compared to approximately five times more for Q$\beta$. Estimated values of $\alpha$, together with their fitting uncertainties, are reported in Table~\ref{tab:populations}. We observe a significant difference between the two systems: SARS-CoV-2 exhibits lower exponent values, typically in the range 1.7--2.0, whereas Q$\beta$ shows consistently higher values, between 2.2 and 2.6. To quantify this difference, we compared the distributions of estimated exponent values using a Welch two-sample $t$-test, obtaining a highly significant separation between the two viruses ($t \approx 5.7$, $p \approx 3 \times 10^{-4}$), with Q$\beta$ exhibiting systematically larger $\alpha$ values than SARS-CoV-2.

These differences in $\alpha$ reflect quantitative changes in the reconstructed genotype networks. The lower exponent observed in SARS-CoV-2 is consistent with a more homogeneous connectivity distribution, particularly among nodes of intermediate degree, indicating a relatively smoother decay in connectivity heterogeneity. By contrast, the larger exponent found in Q$\beta$, together with its broader scaling regime and more gradual decay across the range of $k$ values, is indicative of a more heterogeneous connectivity structure. In this regime, the population organization is more strongly hierarchical, with a clearer distinction between highly connected hubs and the bulk of low- and intermediate-degree nodes. The fast decaying cut-off of SARS-CoV-2 degree distributions reveals a depletion in the degree of the highest connected nodes that is consistent with differences in sequencing coverage.  

Assortativity, quantified through the average nearest-neighbour degree $k_{nn}(k)$ as a function of $k$, shows an inverse dependence over a broad range of $k$ values in all analysed networks, consistent with a scaling behaviour $k_{nn}(k) \propto k^{-\beta}$ (see Fig.~\ref{fig:macro_pnas}b), with $\beta > 0$ indicating a decreasing function of degree. This behaviour is further captured by the assortativity coefficient $-1 \le r \le 1$, defined as the Pearson correlation between the degrees of connected node pairs \cite{newman:2002}. Both measures, $\beta > 0$ and $r < 0$, indicate that the reconstructed genotype networks are strongly disassortative. In other words, low-degree mutants preferentially link to highly connected haplotypes. However, the distribution in Fig.\ \ref{fig:macro_pnas}(b) presents an up-kick for large degree, indicating that hubs also connect to other high-degree nodes. Our genotype networks are strongly disassortative except for this characteristic. These structural features, together with the fat-tailed degree distributions, is consistent with the hierarchical organization observed in Fig.~\ref{fig:lcc_nets}.

Beyond this global property, we observe a systematic relationship between assortativity and sequence coverage $A$. Specifically, the fitted values of $\beta$ show a strong negative correlation with $A$ across all datasets (Pearson correlation $r \approx -0.88$, $p \approx 0.001$), indicating that more deeply sampled populations tend to exhibit smaller $\beta$ values, i.e. weaker degree decay in the nearest-neighbour function. This trend is consistent with previous interpretations \cite{seoane:2026}: as quasispecies diversity increases, the number of intermediate- and high-degree nodes grows, increasing the probability of connections among them and thus reducing the effective disassortativity measured through $\beta$. These results therefore suggest that the change in disassortativity could be a sampling effect. 

The number of triangles per haplotype in Q$\beta$ and SARS-CoV-2 populations shows a pattern consistent with previous observations in other Q$\beta$ populations \cite{seoane:2026}: for large values of the degree $k$, mutations tend to be randomly distributed along the haplotype, and $\triangle(k) \simeq \triangle_{\mathrm{rnd}}(k)$ (see Fig.~\ref{fig:macro_pnas}c,d). In contrast, at low degree values, significant deviations from site neutrality are observed, with an excess of low-degree haplotypes exhibiting values of $\triangle(i)$ more than two standard deviations $\sigma_{\triangle_{\mathrm{rnd}}}$ above the expected random value. No significant deviations towards a deficit of triangles are observed. This excess of triangles at low degree suggests the presence of local constraints in sequence space, where low-connectivity haplotypes form neighbourhoods more clustered than expected under a random mutational model---i.e., there are certain positions that admit more mutations than others.

\subsection{Hierarchical organization around the root haplotype}

Topological measures indicate that all our viral populations are organized in a broadly similar manner. The degree distribution and the disassortative nature of the reconstructed networks, together with representations such as those in Fig.~\ref{fig:lcc_nets}, support a hierarchical organization of extant variation and suggest that additional features of the evolutionary process may be revealed by quantities that depend on the distance $d$ to the most abundant haplotype.

\begin{figure}[H]
    \begin{center}
        \includegraphics[width=0.55\textwidth]{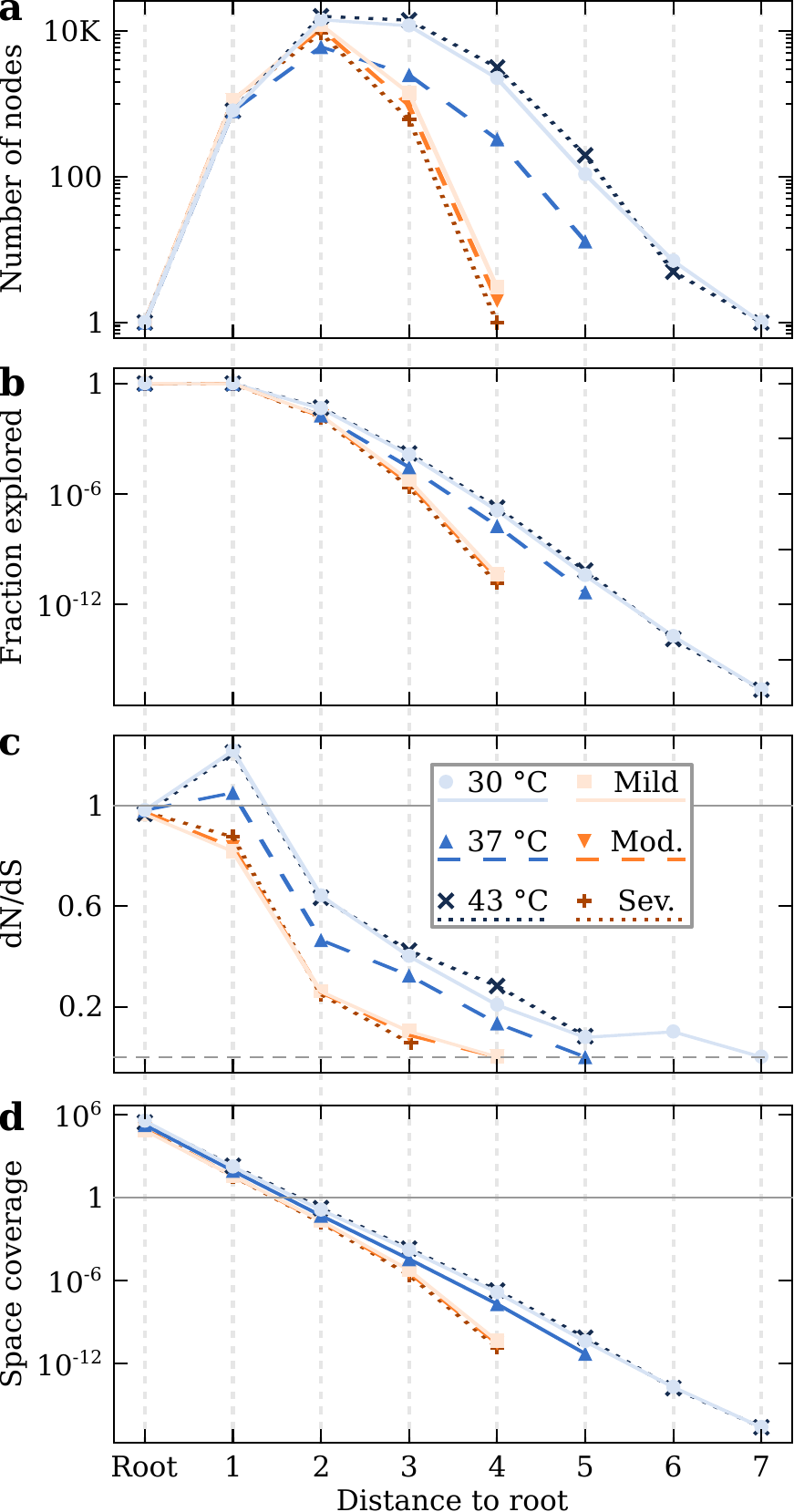}
        \caption{\textbf{Structure of intra-host genotype networks as a function of distance to the root.} Different measures are comparatively explored for the two viruses as a function of Hamming distance to the root sequence (distance level 0): (a) absolute number of nodes (haplotypes); (b) fraction of haplotypes explored; (c) average $dN/dS$ ratio; and (d) average abundance per possible haplotype.}
        \label{fig:micro_pnas}
    \end{center}
\end{figure}

Figure~\ref{fig:micro_pnas} represents four quantities of interest measured as a function of the Hamming distance $d$ (number of mutations) to the most abundant haplotype (dubbed the root) in each of the six samples. All panels illustrate the exhaustive search performed at distances $d=1$ and $d=2$, where almost all possible mutants are sampled at each time step. Since a natural population is 2 to 3 orders of magnitude larger than the size of the deep-sequenced sample, it is to be expected that most mutants until distance $d=3$ are actually present at each passage. Through successive replication cycles and passages, a large fraction of mutants at distance $d=4$ are likely generated. The exploration of mutants at larger distances from the root decays rapidly as a result of the exponential increase in the number of different haplotypes (Eq.\ (\ref{eq:Sd})), together with the decrease in average abundance. Overall, this rapid decay highlights the predominantly local nature of intra-host exploration in genotype space, where evolutionary dynamics efficiently and exhaustively sample the immediate mutational neighbourhood but face progressively stronger constraints in accessing more distant regions of the sequence landscape \cite{manrubia:2026}.

The $dN/dS$ ratio (Fig.\ \ref{fig:micro_pnas}(c)) reflects the balance between synonymous (S) and non-synonymous (N) mutations as a function of distance from the root. This quantity does not account for the abundance of each variant type, but it shows that both synonymous and non-synonymous mutations are exhaustively sampled at $d=1$, where the high abundance of the root ensures near-complete coverage of its mutational neighbourhood, largely independent of the fitness effects of first-step mutants. At larger distances, the $dN/dS$ ratio decreases, and in regions of low abundance the effects of purifying selection become apparent as a relative enrichment of synonymous mutations.

This pattern indicates a progressive shift in the balance between mutation and selection across the network. Close to the root, where mutation supply saturates the local neighbourhood, both types of mutations are frequently generated and observed. At larger distances, however, purifying selection increasingly constrains the persistence of non-synonymous variants, leading to a relative depletion of amino-acid-altering mutations. In this regime, selection becomes detectable because mutation supply is no longer sufficient to uniformly populate the available sequence space.

These results are consistent with the relationship between degree and triangle abundance: at high degree (corresponding to low distance $d$ from the root), no systematic deviation is observed in either the distribution of triangles or the $dN/dS$ ratio, and both quantities are consistent with random assignment of mutations along the haplotype. At low degree, the $dN/dS$ ratio decreases, indicating a relative enrichment of synonymous mutations, while the number of triangles simultaneously shows significant positive deviations from the random expectation. This establishes a correspondence between a purely topological measure (triangle abundance) and a population-genetic quantity (the ratio of non-synonymous to synonymous mutations). Note that the number of triangles is not affected by the bias discussed above that affects Eq.\ \ref{eq:dNdS}. 

Finally, the sequence space coverage $C(d)$ (Fig.~\ref{fig:micro_pnas}d) decays approximately exponentially for all samples, following $C(d) = A(0)\exp\{-d/\delta\}$, where $A(0)$ is the abundance at the root and $\delta$ is a characteristic decay length. The values of $\delta$ are of the same order of magnitude across populations (Table~\ref{tab:populations}), although a Welch two-sample $t$-test indicates a statistically detectable difference between viruses ($t \approx 8.9$, $p < 10^{-5}$), with Q$\beta$ exhibiting slightly larger values. Importantly, $\delta$ shows no significant dependence on sequence coverage, $A$, within either viral group. This suggests that this decay scale reflects intrinsic features of local exploration in genotype space rather than sampling depth.

\section{Effect of sequencing coverage on inferred quasispecies structure}

To assess whether the observed differences between Q$\beta$ and SARS-CoV-2 reflect intrinsic properties of the underlying quasispecies or are instead driven by sequencing depth, we perform a controlled subsampling analysis. The working hypothesis is that several topological quantities inferred from genotype networks are sensitive to coverage, and that differences between viral systems may be partially reproduced by reducing sampling depth in a high-coverage dataset.

Empirically, several observables show a systematic dependence on sequence coverage. For instance, sequence space coverage $C(d)$ exhibits a similar functional form across viruses (Fig.~\ref{fig:micro_pnas}d), although the maximum explored distance is reduced in SARS-CoV-2. Analogous trends are observed for the total number of haplotypes at each distance (Fig.~\ref{fig:micro_pnas}a) and for the fraction of diversity explored (Fig.~\ref{fig:micro_pnas}b). Consistently, the total diversity $D$ is lower in SARS-CoV-2 samples (Table~\ref{tab:populations}), and these quantities correlate with the total number of sequences $A$.

To test whether these patterns can be explained by sampling effects alone, we use a high-coverage Q$\beta$ population (30$^\circ$C) as a reference system and generate artificial datasets by random subsampling of sequences. This procedure preserves the empirical abundance distribution while systematically varying the effective coverage $A_s$, thereby emulating reduced sequencing depth under controlled conditions.

\begin{figure}[H]
    \centering
    \includegraphics[width=0.55\textwidth]{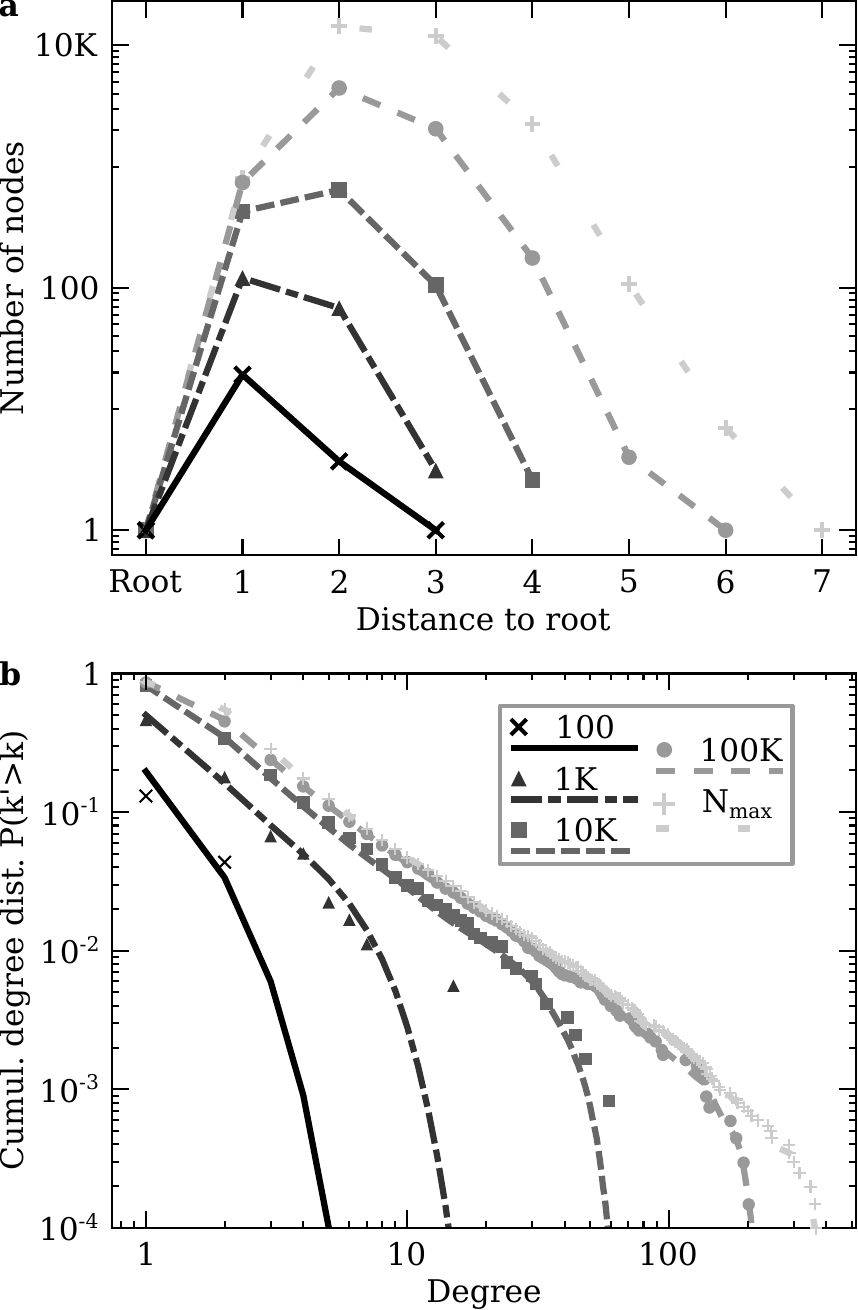}
    % Me di cuenta repasando scripts que la media no estaba bien calculada para todas las distancias porque no estabamos incluyendo todas las reps en el calculo. Me explico: Tomando como ejemplo la el sampling size = 100 vemos que para d=3 la media es 1, esto implicaria que en las 100 reps hemos encontrado 1 haplotipo a esa distancia. Pero eso no era cierto, teniamos 7 out of 100 reps en los que teniamos 1 haplotipo pero la media tendría que ser calculada añadiendo 93 ceros, que es el numero de veces que no llegamos al nivel 3 en el sampleo. De esa manera, las medias pasaran a estar por debajo de 0 en todos esos casos. Lo dejo anotado para que no se me olvide que tengo que cambiarla y por si hubiera que actualizar algún comentario. 
    \caption{\textbf{Effect of sequencing coverage on inferred genotype network structure.} (a) Complementary cumulative degree distribution (CCDF) for varying sequence coverage. (b) Number of nodes (haplotypes) as a function of Hamming distance from the root sequence (distance level 0) for different levels of sequence coverage. Data are obtained by random subsampling of the Q$\beta$ (amplicon 1) population at 30$^\circ$C. If sample size is smaller than $A_{max}$, each curve corresponds to averages over 100 independent realizations of the subsampling and network reconstruction procedure.}
    \label{fig:sampling_nS_nH}
\end{figure}

Figure~\ref{fig:sampling_nS_nH}a shows the complementary cumulative degree distributions (CCDFs) obtained under the same subsampling scheme. The results demonstrate a systematic modification of the degree distribution at low coverage, where the apparent power-law regime is progressively shortened, developing a cut-off consistent with a reduction of the degree in high-degree nodes. For values of $A_s$ comparable to those of SARS-CoV-2 datasets, the resulting CCDF closely reproduces the empirical curves in Fig.~\ref{fig:macro_pnas}a, indicating that differences in the inferred tail behaviour can arise naturally from incomplete sampling.

Figure~\ref{fig:sampling_nS_nH}b shows the number of haplotypes detected at each distance $d$ for different subsample sizes $A_s$. For each $A_s$, sequences are drawn at random according to their observed abundances, and the corresponding largest connected component is reconstructed as described above. Each curve corresponds to an average over 100 independent realizations of the subsampling and reconstruction procedure. As coverage decreases, the inferred exploration of sequence space becomes increasingly restricted, reproducing the qualitative behaviour observed in SARS-CoV-2 datasets of lower coverage (Fig.~\ref{fig:micro_pnas}a).

Overall, these results support the interpretation that several apparent differences between viral populations are, at least in part, attributable to coverage effects rather than fundamental differences in quasispecies organization. In particular, subsampling reproduces key trends in diversity, connectivity, and degree distributions, highlighting the importance of accounting for sequencing depth when comparing genotype network structure across datasets.

\section{Discussion}

This study compares the structure of genotype networks in two RNA viruses evolving under fundamentally different conditions: bacteriophage Q$\beta$ in controlled laboratory environments and SARS-CoV-2 within infected human hosts. The hierarchical structure of viral quasispecies, previously described for Q$\beta$ \cite{seoane:2026}, is here shown to extend to SARS-CoV-2 populations evolving {\it in vivo}. Despite large differences in genome size, mutation rate, and ecological context, both systems display a remarkably similar organization of their mutant spectra. Importantly, the consistency of all main topological measures across independent amplicons (see Supplementary Information) supports the robustness of the inferred network properties and indicates that the reported differences between SARS-CoV-2 and Q$\beta$ are not driven by local genomic context.

Measures broadly applied to characterize complex networks \cite{newman:2010}, such as the degree distribution and the number of triangles (a measure of clustering), are quantitatively comparable across all populations analysed for both viruses. This consistency further supports the existence of a shared network architecture underlying the mutant spectra of Q$\beta$ and SARS-CoV-2 Populations show some variation in their assortativity, although this quantity is expected to depend on the fraction of the underlying neutral network explored (hence on population diversity) \cite{seoane:2026}, and wide variation has been reported in genotype networks of different origins \cite{aguirre:2011,aguilar:2018,williams:2022}. 

The organization of these viral populations is characterized by a hierarchical genotype network centered around a dominant haplotype and surrounded by layers of variants at increasing Hamming distance. The dense occupancy observed in the immediate neighborhood of the dominant haplotype is consistent with near-exhaustive local exploration of sequence space. Given the high mutation supply in RNA viruses, most single-mutant variants are expected to be generated repeatedly \cite{domingo:2012,lauring:2010,sender:2021}, implying that the presence of a variant does not necessarily indicate neutrality or selective advantage, but may instead reflect the stochastic generation of mutations in highly abundant genetic backgrounds.

%The similarity across sampled populations is particularly striking given the lower mutation rate of SARS-CoV-2. 
Apparent differences between the two viruses can be largely attributed to sampling effects, as supported by subsampling experiments in Q$\beta$: lowering coverage yields sequence ensembles that reproduce the topological and hierarchical features observed in SARS-CoV-2 datasets. The emergence of comparable network architectures suggests that the global organization of quasispecies is neither fundamentally determined by the absolute number of mutations per genome, nor by environmental differences, but by more general constraints governing the exploration of sequence space. 

Beyond minor quantitative differences, the internal organization of genotype networks appears highly conserved and can be understood as the outcome of three general processes. First, mutation continuously generates a dense cloud of single-mutant neighbors around the dominant haplotype through a process akin to diffusion. Second, the number of possible genotypes increases exponentially with Hamming distance, leading to progressively sparser sampling of sequence space. This fact notwithstanding, the local sampling is so exhaustive that we estimate that all mutants up to distance $3$ are likely present in the complete quasispecies. Third, purifying selection increasingly limits the persistence of variants carrying multiple mutations. Together, these processes naturally produce a hierarchical structure and a highly heterogeneous exploration of sequence space, a mechanism that may represent a generic feature of RNA virus populations and, more generally, of ensembles of replicators \cite{manrubia:2026}.

This distance-dependent regime reconciles two key observations. In the immediate neighborhood of the dominant haplotype, mutational space is densely and nearly exhaustively explored, and even deleterious mutations can be repeatedly generated and transiently observed. At larger mutational distances, exploration becomes progressively sparser and selection increasingly determines which variants persist. This shift in the balance between mutation and selection is consistent with the observed decay in $dN/dS$ with distance, as well as with changes in local network structure (noted by $\triangle(i) > \triangle_{rnd}$).

These results support a view of viral evolution as a continuous, out-of-equilibrium process in which populations expand locally in sequence space through mutation-driven diffusion, while selection shapes the distribution of variant abundances. In this framework, exploration does not rely primarily on neutral pathways but emerges from the high connectivity of genotype space and the constant generation of variants in the immediate neighborhood of highly abundant sequences. 

The hierarchical organization of genotype networks and their persistence, even in samples where mutations have recently been fixed, also has important implications for viral transmission and adaptation. Transmission bottlenecks are expected to sample only a small subset of the mutant spectrum, typically variants close to the dominant haplotype. However, once replication resumes in the new host, the high mutation rate characteristic of RNA viruses should rapidly regenerate a cloud of mutants around the transmitted genome. Our results strongly point at hierarchical genotype networks acting as structural attractors, that are repeatedly reconstructed during viral evolution, with selective pressures determining which haplotype occupies the central position while the overall geometry of the mutant cloud remains largely invariant.

More generally, genotype networks capture aspects of viral population structure that are not evident from diversity measures alone. While previous studies have reported differences in mutant spectrum complexity across clinical categories \cite{martinezgonzalez:2022}, such differences do not necessarily translate into qualitative changes in global network organization. Rather, diversity and network structure provide complementary information: the former reflects the distribution of variant frequencies, whereas the latter encodes their mutational connectivity and is comparatively robust to sampling effects.

Taken together, the existence of a shared genotype network architecture across viruses with distinct life histories points to quasispecies organization being governed by general principles. This convergence is particularly striking given the profound differences in the evolutionary processes shaping each system. In the case of SARS-CoV-2, viral populations evolve within individual hosts under heterogeneous and fluctuating selective pressures, including innate and adaptive immune responses, tissue compartmentalization, and genetic backgrounds shaped by prior transmission events. By contrast, Q$\beta$ evolves in a highly controlled {\it in vitro} setting, where adaptation proceeds through serial passages imposing recurrent population bottlenecks and where the dominant selective pressure is externally defined. Despite these contrasting demographic, ecological, and selective regimes, both viruses give rise to genotype networks with the same hierarchical architecture. This observation suggests that while selective pressures determine which haplotype becomes dominant in a given environment, they do not substantially alter the overall organization of the mutational neighborhood that emerges around that haplotype. Testing this hypothesis along different SARS-CoV-2 waves, across additional RNA viruses, and extending it to DNA viruses with lower mutation rates, will help determine the extent to which hierarchical genotype networks represent a universal organizing principle of viral evolution.

%Taken together, the existence of a shared genotype network architecture across viruses with distinct life histories suggests that quasispecies organization may be governed by general principles. Testing this hypothesis with data from additional SARS-CoV-2 waves, across additional RNA viruses, and extending it to DNA viruses with lower mutation rates, will help determine the extent to which hierarchical genotype networks represent a universal organizing principle of viral evolution.

%%%%%%%%% IMPORTANT COMMENT HERE %%%%%%%%%
% Dejo un fichero llamado ./Figures/S_A2-reg2/Labels_summary.xlsx con los datos referentes a cada poblacion con datos secuenciados para el amplicón 2 de Qbeta que son los que empleamos en las tres últimas figuras. En resumen, las 3 muestras seleccionadas mantienen su temperatura constante hasta el pase seleccionado. No obsante, en el caso de 40 grados el experimento es distinto. No es un experimento de temp constante durante 60 pases como en el caso de 30 y 43 sino que es un experimento de 30 pases a 40 grados y 30 pases a 43. Todos los pases hasta el 30 (incluido) han mantenido la temp constante y por eso podemos emplearlos. No tenemos ningún pase a 37 grados con el amplicon 2 secuenciado. Por eso cogemos uno de 40. En cuanto al ancestro, todas mantienen el P2 como en el caso del amplicón 1 salvo las de 40 grados también que emplean P25. 
%%%%%%%%% IMPORTANT COMMENT HERE %%%%%%%%%

\section{Acknowledgments}
This research was supported through grants PID2023-153225NA-I00 (L. F. S.), PID2023-146622OB-I00 (B. M.-G. and C. P.), PID2023-147963NB-C21 (S. M.-A., I. A. and S. M.) and PID2023-147963NB-C22 (P. S. and E. L.) funded by MICIU/AEI/10.13039/501100011033 and by European Regional Development Fund, European Union. L. F. S.  received funding from the Occident Foundation (Grant No. FJSCNB-2022-12-B). B. M.-G. and C. P. were supported  through the 'Centro de Excelencia Severo Ochoa' award to the Centro de Biología Molecular Severo Ochoa (CBM) (Grant No. CEX2021-001154-S); project TEC-2024/BIO-66 (SALAINDEC-CM from Comunidad de Madrid/FEDER), and the European Commission-Next Generation EU (regulation EU 2020/2024) through the Consejo Superior de Investigaciones Científicas (CSIC’s) Global Health Platform [Plataforma Temática Interdisciplinar (PTI) Salud Global]. Institutional grant from the Fundación Ramón Areces Foundation to the CBM is also acknowledged.

%\bibliographystyle{plain}
%\bibliography{bibliography}
\printbibliography

\newpage

\section*{Supplementary Information for \\
Shared quasispecies architecture in experimental and natural RNA virus populations}

The results presented in the main manuscript analysed and compared the structure of genotype networks for a single amplicon of the two viruses. To assess the robustness of the reported results with respect to genome location, we repeat the full analysis for a second, independent amplicon in Q$\beta$ and SARS-CoV-2, as described in Section~2. This allows us to test whether the observed hierarchical organization and associated topological properties depend on the specific genomic fragment or instead reflect general features of the underlying quasispecies structure.

The Q$\beta$ samples used for this analysis in the main ms and in this SI correspond to independent adaptation experiments at 30$^\circ$C, 37$^\circ$C, and 43$^\circ$C. These populations were propagated under constant temperature conditions up to the passage considered for sequencing, with the exception of the 40$^\circ$C condition, which corresponds to a mixed-temperature evolution protocol (30 passages at 40$^\circ$C followed by 30 passages at 43$^\circ$C). Accordingly, only passages prior to the temperature shift are included in the present analysis, ensuring consistency in the evolutionary regime across samples. In addition, while most Q$\beta$ populations originate from the same ancestral population (P25, corresponding to an extensively pre-adapted ancestor), the 40$^\circ$C lineage originates from a lineage passaged twice under laboratory conditions (P2). This difference is explicitly noted when interpreting results but does not affect the structural comparisons reported here. 

\begin{figure}[t]
    \begin{center}
        \includegraphics[width=\textwidth]{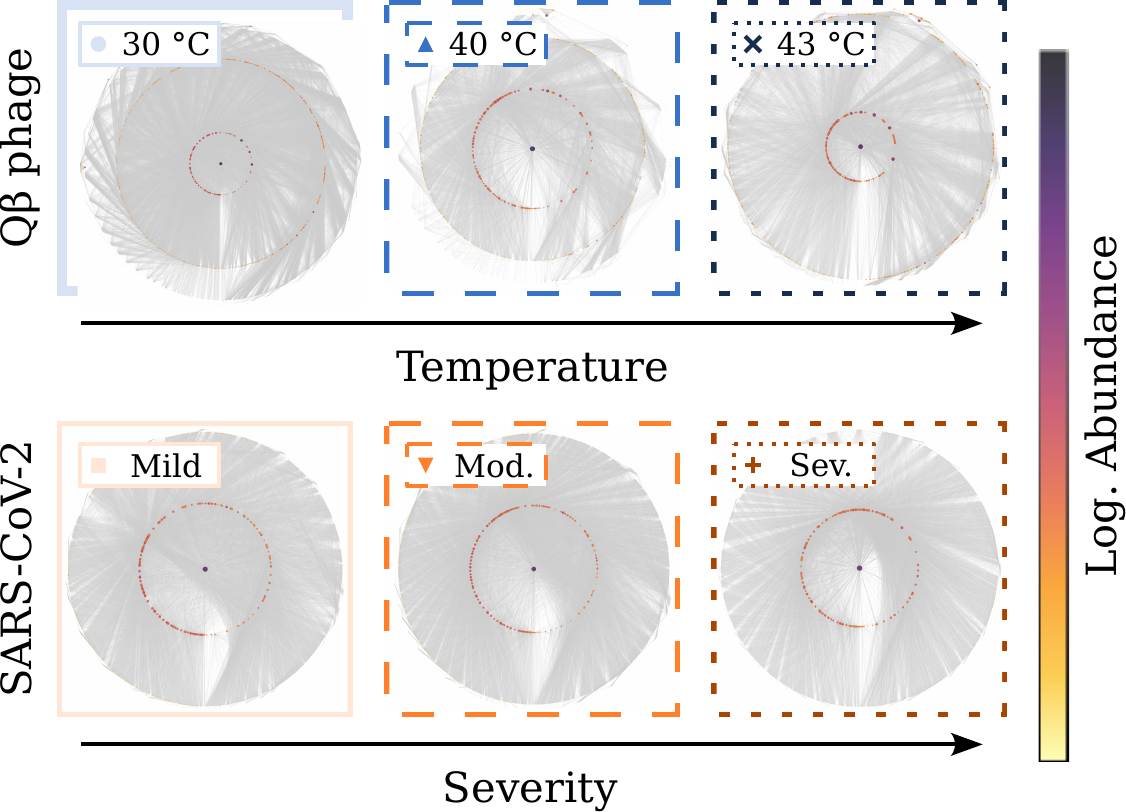}
        \caption{\textbf{Hierarchical representation of genotype networks for an independent amplicon.} Largest connected components of genotype networks for amplicon 2 of SARS-CoV-2 and Q$\beta$ (see Materials and Methods). Node size and node color represent haplotype abundance. The same hierarchical organization observed in the main text is preserved across independent genomic regions.}
        \label{fig:lcc_nets_S_A2}
    \end{center}
\end{figure}

\begin{figure}[t]
    \begin{center}
        \includegraphics[width=\textwidth]{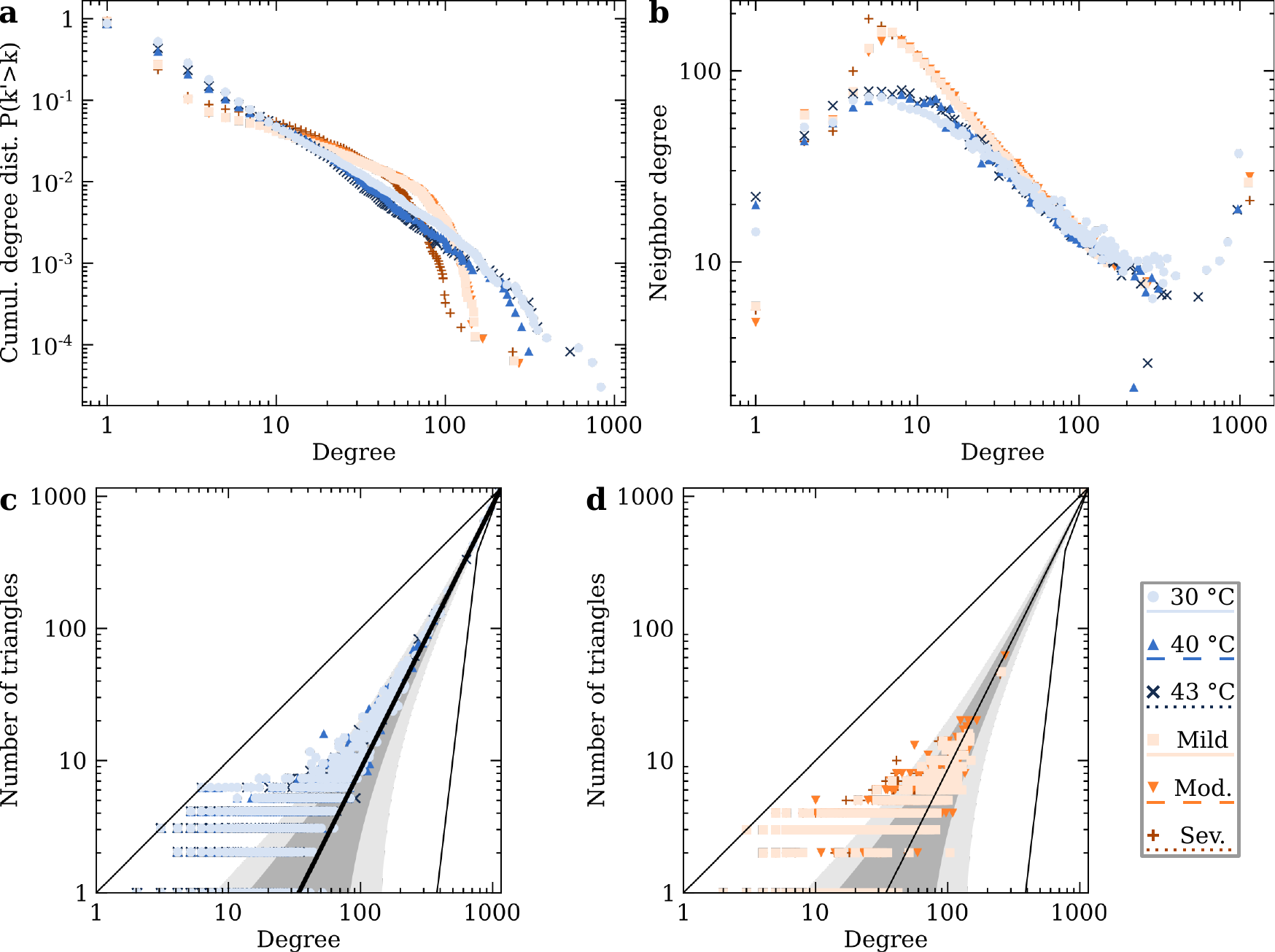}
        \caption{\textbf{Topological structure of the LCC of genotype networks for an independent amplicon.} (a) Complementary cumulative degree distribution (CCDF). (b) Assortativity as a function of degree. (c,d) Number of triangles per haplotype. In (c,d), black lines are bounds to the expected number of triangles as a function of node degree (upper and lower lines) and the bold black line represents expected values if mutations are randomly assigned along the haplotype sequence; dark gray and lighter gray regions correspond to one and two standard deviations from average, as in Fig.~3. The same qualitative patterns observed in the main text are recovered across both viruses and amplicons.}
        \label{fig:micro_pnas_S_A2}
    \end{center}
\end{figure}

\begin{figure}[t]
    \begin{center}
        \includegraphics[width=0.55\textwidth]{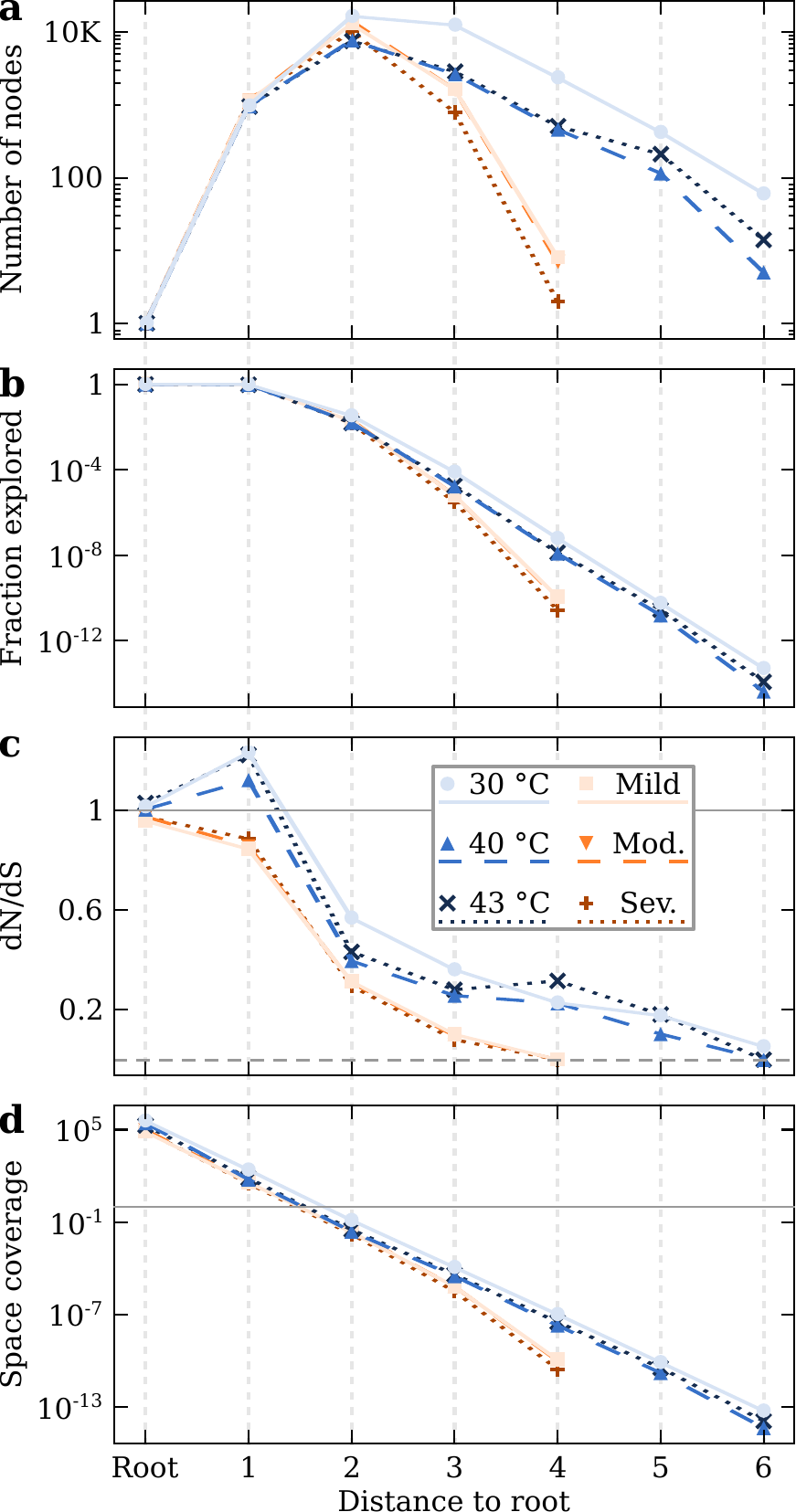}
        \caption{\textbf{Structure of intra-host genotype networks as a function of distance to the root for an independent amplicon.} (a) Number of haplotypes as a function of Hamming distance from the root sequence. (b) Fraction of explored haplotypes. (c) Average $dN/dS$ ratio. (d) Average abundance per possible haplotype. All quantities reproduce the distance-dependent trends reported in the main text (Fig.~4).}
        \label{fig:macro_pnas_S_A2}
    \end{center}
\end{figure}

Figures~\ref{fig:lcc_nets_S_A2}--\ref{fig:macro_pnas_S_A2} show the corresponding results for amplicon 2 of Q$\beta$ and SARS-CoV-2. Overall, the same qualitative patterns observed in the main text are recovered. In particular, both amplicons exhibit fat-tailed degree distributions, disassortative connectivity, and a consistent hierarchical organisation of the largest connected component. Likewise, the scaling behaviour of macroscopic quantities and their dependence on distance from the root are preserved, including the decay of sequence space exploration and the structure of local mutational neighbourhoods.

Quantitatively, the fitted parameters reported in Table~1 for the second amplicon are consistent with those obtained for the first amplicon within statistical uncertainty, with only minor variations attributable to differences in sampling depth and local sequence context. Importantly, the relative differences between Q$\beta$ and SARS-CoV-2 are maintained across amplicons, supporting the robustness of the main conclusions regarding differences in exponent values, assortativity, and hierarchical structure.

The analysis of an independent amplicon confirms that the hierarchical organization and the main topological signatures of genotype networks are conserved across genomic regions, indicating that the observed properties are not specific to a particular sequence fragment but reflect general features of the underlying viral quasispecies structure. Indeed, the fitted exponents ($\alpha$, $\beta$, and $\delta$) obtained for the second amplicon fall within the uncertainty ranges of those reported in the main text, confirming quantitative consistency across genomic regions. Importantly, these differences in experimental history do not alter the observed topological invariants of the reconstructed networks.

% plot distribución de grado, triángulos, asortatividad

\end{document}